\documentstyle[twocolumn,prc,aps,floats,epsf]{revtex}

\draft
\begin{document}

\wideabs{

\title{Tests of Transfer Reaction Determinations of
Astrophysical S-Factors}

\author{C.A. Gagliardi, R.E. Tribble, A. Azhari, H.L. Clark,
Y.-W. Lui, A.M. Mukhamedzhanov, A. Sattarov, L. Trache}
\address{Cyclotron Institute, Texas A\&M University,
College Station, 
Texas 77843} 
\author{V. Burjan, J. Cejpek, V. Kroha, \v{S}. Pisko\v{r}, J.
Vincour}
\address{Institute for Nuclear Physics, Czech Academy of
Sciences,
Prague-\v{R}e\v{z}, Czech Republic}  
\date{\today}
\maketitle

\begin{abstract}
The ${}^{16}O ({}^{3}He,d) {}^{17}F$ reaction has been
used to determine
asymptotic normalization coefficients for transitions to the
ground and first
excited states of ${}^{17}F$.  The coefficients provide the
normalization for the
tails of the overlap functions for ${}^{17}F \to{}^{16}O + p$
and allow us to
calculate the S-factors for ${}^{16}O (p,\gamma){}^{17}F$ at
astrophysical energies. The calculated S-factors
are compared to measurements and found to be in very good
agreement.  This provides the first test of this indirect
method to determine
astrophysical direct capture rates using transfer reactions.
In addition, our results yield S(0) for capture to the ground and
first excited states in $^{17}F$, without the uncertainty associated
with extrapolation from higher energies.
\vskip 5mm
To be published in Phys. Rev. C.
\end{abstract}

} 

\section{Introduction}

Nuclear capture reactions, such as $(p,\gamma)$ and
$(\alpha,\gamma)$, play a major role in
defining our universe.
A primary goal in nuclear astrophysics is to
determine rates for capture reactions that are important
in the evolution of stellar
systems.  However, the reactions of interest often involve
radioactive targets which makes
measurements quite difficult or even impossible using
conventional methods.  Hence techniques have been developed to
determine rates by indirect methods.  For example, precise
information
on excitation energies and particle decay widths can be used to make
accurate predictions of rates which proceed by resonance capture. 
The
only reliable method to determine a reaction rate that is dominated
by
direct capture has been to measure it at laboratory energies with a
low energy particle beam and then extrapolate the result to energies
of astrophysical interest.

Attempts have been made
to use both Coulomb dissociation
\cite{Moto94} and the determination of asymptotic normalization
coefficients (ANC) from conventional nuclear transfer reactions
\cite{xu94,Liu96,Gagl98,Vanc98} to determine
S-factors for direct capture reactions, but neither technique has
been tested to verify its reliability.  Such tests are
crucial, as stressed in the most recent evaluation of
solar fusion cross section rates \cite{rmp98}.  We report here
the first test of one of these two techniques to determine
astrophysical S-factors; we demonstrate that the ANC inferred
from a measurement of a proton transfer reaction
can directly determine a $(p,\gamma)$ direct capture rate at
astrophysical energies.

Direct capture reactions of astrophysical
interest usually involve systems where the binding energy
of the captured proton
is low.  Hence at stellar energies, the capture proceeds
through the tail of the
nuclear overlap wave function.  The shape of this tail is
completely determined by the Coulomb interaction, so the rate of the
capture reaction can be calculated accurately if one knows its
amplitude.  The
asymptotic normalization coefficient $C$ for the system
$B \leftrightarrow A+p$ specifies the amplitude of the
single-proton tail of the wave function for nucleus $B$ when the
core $A$ and the proton are separated by a distance large
compared to the nuclear radius.  Thus, this
normalization coefficient determines the corresponding direct
capture rate.

The advantage of the ANC approach is that it provides a method
to determine direct capture S-factors
accurately from the results of nuclear
reactions such as peripheral nucleon transfer which can be
studied with radioactive beams and have cross
sections that are orders of magnitude larger than the direct
capture reactions themselves.  Furthermore, direct capture S-factors
derived with this technique are most reliable at the
lowest incident energies,
precisely where capture cross sections are smallest and most
difficult
to measure directly.  In fact, the ANC approach even permits one to
determine S-factors at zero energy, which is not possible with direct
measurements except by extrapolation.

While there is little
controversy that knowledge of the {\it asymptotic normalization
coefficient} for a loosely bound
nuclear system allows one to compute the corresponding direct
capture rate, the nuclear astrophysics community has clearly
indicated \cite{rmp98} that a test of the relationship between
the {\it transfer reaction cross section} and the astrophysical
S-factor is important to validate this approach.  The community's
skepticism originates in the well-known model dependence found in
distorted-wave Born approximation
(DWBA) analyses of transfer reaction data in terms of
spectroscopic factors, which is due to the uncertainty in the
DWBA calculations associated with the choice of optical model
potentials and single particle wave functions.  By parametrizing
the DWBA cross section of a peripheral transfer reaction in terms
of ANC's, rather than spectroscopic factors, we can reduce the
uncertainty associated with the choice of single
particle wave functions so that it becomes small compared to that
associated with the optical potential \cite{mukh98,Trac98}.  By
choosing
appropriate reactions, beam energies and scattering angles, we
can also minimize
the uncertainty associated with the choice of optical
model potentials.

In this article, we describe a measurement of the
${}^{16}O({}^{3}He,d){}^{17}F$ reaction, from which we
determine the ANC's for
the ${\frac{5}{2}}^+$ ground state and the ${\frac{1}{2}}^+$ first
excited state in ${}^{17}F$.
We then use our measured ANC's to calculate, with no additional
normalization factors, the
S-factors for the ${}^{16}O(p,\gamma){}^{17}F$
reaction at astrophysical energies.  Such a determination of the
S-factors
for ${}^{16}O (p,\gamma) {}^{17}F$ from
its ANC's measured in proton transfer reactions is an ideal test
case for this indirect method \cite{rmp98} because the
results can be compared to existing direct measurements
of the capture cross sections \cite{capt1,capt2}.
Furthermore, the ${}^{16}O (p,\gamma) {}^{17}F$
reaction has substantial similarities to the
$^7Be(p,\gamma){}^8B$ reaction, which is the
source of all high energy neutrinos produced in the sun.
We will report determinations of the S-factor for
$^7Be(p,\gamma){}^8B$ using this
technique in future publications.  It will also be
straightforward to utilize this procedure to
determine S-factors
at astrophysical energies for other cases that include
significant direct capture components.

\section{$^{17}F \leftrightarrow {}^{16}O + \lowercase{p}$ Asymptotic
Normalization Coefficients}

For a peripheral transfer reaction, ANC's are extracted
from the measured angular
distribution by comparison to a DWBA calculation.
Consider the proton transfer
reaction $a + A \to c + B$,
where $a=c + p$ and $B= A + p$.  The experimental
cross section is related to the DWBA calculation according to 
\begin{equation}
\frac{d\sigma}{d\Omega}=
\sum_{l_B j_B l_a j_a} (C^{B}_{A p l_B j_B})^{2}
(C^{a}_{c p l_a j_a})^{2} R_{l_B j_B l_a j_a} ,
\label{dwcs2}
\end{equation}           
where 
\begin{equation}
R_{l_B j_B l_a j_a}=\frac{{\tilde \sigma}_{l_B j_B l_a j_a}}
{b^{2}_{A p l_B j_B} b^{2}_{c p l_a j_a}} .
\label{r1}
\end{equation}
${\tilde \sigma}$ is the calculated DWBA cross section
and the $b$'s are the asymptotic  normalization
constants for the single particle orbitals used in the
DWBA.  The sum in
Eq.~(\ref{dwcs2}) is taken over the allowed angular
momentum couplings, and the
$C$'s are the ANC's for $B \to A + p$ and $a \to c + p$.
The normalization of the DWBA cross section by the
ANC's for the single
particle orbitals makes the extraction of the ANC for
$B \to A + p$
insensitive to the parameters used in the
single particle potential wells \cite{mukh98,Trac98}, in contrast to
traditional spectroscopic factors.
See \cite{mukh98} for additional details.  

DWBA calculations of the ${}^{16}O ({}^{3}He,d) {}^{17}F$
reaction populating the $^{17}F$ first excited state indicate
that the sensitivity of the extracted ANC to the choice of
optical model potentials is minimized near $0^{\circ}$.
There exists a previous study of the
${}^{16}O ({}^{3}He,d) {}^{17}F$
reaction at $E_{{}^{3}He}$ = 25 MeV \cite{Vern94} that
reported cross sections at 9 angles
over the range $\theta_{cm} \approx 6-36^{\circ}$. 
The limited small-angle coverage makes an attempt to infer the
$^{17}F$ first excited state ANC from that experiment very imprecise. 
We have now measured the
${}^{16}O ({}^{3}He,d) {}^{17}F$ reaction at $E_{{}^{3}He}$ $\approx$
29.7 MeV primarily to determine the angular distribution
carefully at small angles, thus minimizing the systematic
uncertainty in the extracted ANC.  However, by obtaining data at a
second beam energy, we can also do a combined analysis
to reduce our sensitivity to the choice of optical
potentials even further.

Two separate measurements were performed, one optimized to
determine the absolute cross section with a minimum of
uncertainty and the other to obtain a detailed angular
distribution at small angles.  The reaction was measured at
laboratory angles between $6.5^{\circ}$ and
$25^{\circ}$ using a momentum-analyzed 29.75 MeV ${}^{3}He$ beam from
the U-120M isochronous
cyclotron of the Nuclear Physics
Institute (NPI) of the Czech
Academy of Sciences incident on a Mylar target.  The target thickness
was measured to be 134
$\mu$g/cm$^2$ by scanning with well-collimated alpha particles
from $^{241}Am$, $^{238}Pu$ and $^{244}Cm$.  Reaction products
were observed by a pair of detector telescopes, consisting of 150
$\mu$m
thick $\Delta E$ and 2000 $\mu$m thick $E$ Si surface barrier
detectors, with solid angles of 0.23 msr.  One of the telescopes was
rotated about the target during the measurements
while the other was fixed at
${\theta}_L = 18.2^{\circ}$. 
Elastic scattering and several reaction
channels were measured
simultaneously in both telescopes to provide a continuous
calibration
of the beam energy, reaction angle and target thickness.  The beam
current was integrated by a Faraday cup biased to 1 kV.
Absolute cross sections were determined to $\pm 4.5\%$, using
procedures developed at NPI to
minimize overall normalization uncertainties \cite{Burj94,Burj96}.

Small angle data at laboratory angles between $1^{\circ}$ and
$11^{\circ}$ were obtained
using a molecular ${({}^{3}He-d)}^+$ beam from the
Texas A\&M University
K500 superconducting cyclotron incident on a 540 $\mu$g/cm$^2$ Mylar
target.  The angular spread of the beam on target was
$\approx$0.1$^{\circ}$ after passing through the Texas A\&M Beam
Analysis System \cite{Youn95}.  Reaction products were detected
at the focal plane of
the Multipole Dipole Multipole magnetic spectrometer \cite{Prin86}
using the modified Oxford detector \cite{Youn95a}.
The detector consists of a 50 cm long gas ionization chamber
to measure the specific energy loss of particles in the gas and their
focal plane positions at four resistive wires, separated by 16 cm
steps
along the particles' trajectories, followed by an NE102A plastic
scintillator to measure the residual energy.  The $^3He$ energy in
the molecular beam was determined from the crossover between the
$^{12}C(^3He,t)^{12}N$ and $^{16}O(^3He,\alpha)^{15}O$ reactions,
observed simultaneously off the Mylar target.  It was 29.71 MeV,
tuned
to match the measurements carried out at the NPI.  The beam angle was
determined to $\pm$0.1$^{\circ}$ from the crossover between the
$^1H(^3He,^3He)^1H$ and $^{12}C(^3He,^3He)^{12}C^{*}$(4.44 MeV)
reactions, also observed simultaneously off the Mylar target.  The
charge in the beam was collected in a
Faraday cup and provided the normalization between different
scattering angles.  The spectrometer has an acceptance of $\Delta
\theta_L = 4^{\circ}$, which was divided into 8 separate
0.5$^{\circ}$
angle bins by ray tracing.  It was moved in 2$^{\circ}$ steps from
laboratory angles of 3$^{\circ}$ to 9$^{\circ}$.  With this
procedure,
the internal consistency of the normalization between angles was
verified.  Additional details regarding the experimental procedures
may be found in \cite{mukh98}.

\begin{figure}[tb]
\begin{center}
  \mbox{\epsfysize=7.5cm \epsfbox{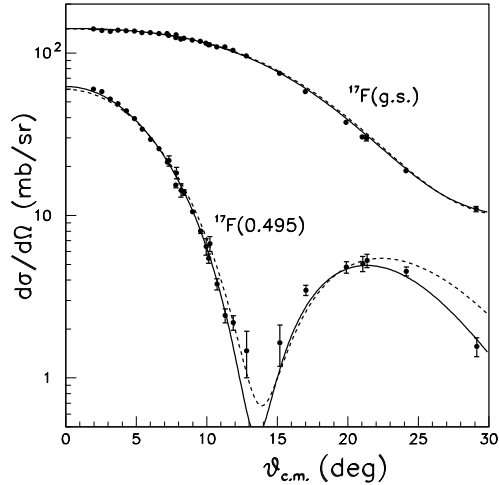}}
\end{center}
\caption{Angular distributions for the ground
and first excited states of
${}^{17}F$ from the
${}^{16}O ({}^{3}He,d) {}^{17}F$ reaction.
The dashed curves
are DWBA fits using optical
potential set I in Table I, and the solid curves
use set II.}
\label{fig1}
\end{figure}

\begin{table*}[tb]
\caption{Adopted optical potentials.  Sets I and II gave
the smallest and largest
ANC's for the two transitions, with other optical potential
combinations giving ANC's in between.  The $d$ potentials
are specified for the
$^{17}F$ first excited state.  Energy-dependent
terms were slightly different for the ground state.
All energies are in MeV, distances
are in fm, and ANC's are in fm$^{-1}$.}

\begin{tabular}{|c|ccc|cccc|ccc|c|cc|}
Set&$V$&$r$&$a$&$W_S$&$W_D$&$r_I$&
$a_I$&
$V_{LS}$&$r_{LS}$&$a_{LS}$&$r_C$&
$C_{d_{5/2}}^2$&$C_{s_{1/2}}^2$\\
\hline
I:~$^3He$&185.03&1.15&0.672&&
11.75&1.511&0.748&&&&1.4&&\\
I:~~$d$&85.87&1.17&0.746&0.60&
12.17&1.325&0.67&6.69&1.07&0.66&1.3&1.00&5980\\
\hline
II:~$^3He$&183.33&1.15&0.659&7.93
&&2.142&0.695&&&&1.4&&\\
II:~~$d$&83.02&1.13&0.80&&12.0
&1.442&0.714&5.2&0.85&0.475&1.3&1.16&7000\\
\end{tabular}
\end{table*}

The absolute normalization
of the Texas A\&M cross section measurements was determined by
matching the ground and
first excited
state yields to those determined at NPI in the angular region where
the two data sets overlap.  The matching procedure introduced an
additional $\pm$1.1\% uncertainty in the absolute normalization of
the
small angle cross section measurements.  The combined angular
distributions for the ground and first excited states
are shown in Fig.~\ref{fig1}.

DWBA calculations were carried out with the
finite range code PTOLEMY
\cite{dwba}, using the full transition operator.
Seven different optical potentials were
studied for the
${}^{3}He - {}^{16}O$ entrance channel.  Six came
from an extensive study of
$^3He$ elastic scattering on $s-d$ shell nuclei at 25
MeV \cite{Vern82}, with small
($<$0.5\%) adjustments in the depths to account for
the energy-dependence
of the real and imaginary volume integrals
\cite{Tros87}.  One came from a global
fit \cite{Tros87}.  The potentials include three
different families of discrete ambiguities,
characterized by the real volume integral, and
contain both volume and surface
imaginary forms.  In general, the calculations with potentials
including volume imaginary terms reproduced our measured angular
distributions slightly better.  Eventually, the potentials with the
intermediate real volume
integrals, which were identified as the ``physical"
family in \cite{Vern82}, were
adopted.  The deep potentials predicted a forward
maximum for the $^{17}F$
excited state that varied too slowly with angle compared to our
measured angular distribution.
Some of the shallow potentials gave reasonable fits
to our measured angular
distributions at 29.7 MeV but did a poor job
reproducing the 25 MeV data
\cite{Vern94}.  Five $d - {}^{17}F$ exit
channel potentials
were studied.  Three came from various global fits
\cite{Daeh80}, and
two came from fits to $d - {}^{17}O$ elastic
scattering \cite{Li76}.
One global potential predicted a forward maximum
that varied too slowly with angle, while the two $d - {}^{17}O$
potentials gave very
poor fits.  The remaining global potentials reproduced the measured
angular distributions well and were adopted.  The single particle
orbitals
were calculated in
Woods-Saxon potentials with $r_0$ in the range
$1.15-1.35$ fm and $a_0$
in the range $0.55-0.75$ fm.  Over this full range, the extracted
$^{17}F$ ANC's varied by only $\pm 1.5\%$ and $\pm 4\%$ for the
ground and first excited states, respectively, demonstrating the
insensitivity of the extracted ANC's to assumptions about the
$^{17}F$
wave functions in the nuclear interior.  In contrast, the more
traditional spectroscopic factors varied by $\pm 45\%$ and $\pm
19\%$.

Normalizing the DWBA calculations
to the data and accounting for the ANC's for the single
particle orbitals and the known
ANC for ${}^{3}He \to d + p$
\cite{Kami90,akram} provides $C^2$
for ${}^{17}F \to{}^{16}O + p$.
Fits over several
angular ranges, from $\theta_{cm}=2-6^{\circ}$ to
$\theta_{cm}=2-30^{\circ}$, gave ANC's consistent
to within $2\%$.  The final ANC's were
determined from fits to the forward angle peaks
($\theta_{cm}=2-9^{\circ}$) to minimize the
sensitivity to the choice of optical model parameters.
Table I shows the adopted optical model parameter
combinations that gave the smallest and largest ANC's.
It is worth noting that most optical potentials
that gave poor fits
nonetheless gave ANC's that also fell within this range.
The corresponding fits to the ground and first excited state
angular
distributions are shown in Fig.~\ref{fig1}.  The fits to the
$^{17}F$ first excited state near the minimum and the weak population
that we observe for the $^{17}F$ ${\frac{1}{2}}^-$ second excited
state and ${\frac{5}{2}}^-$ third excited state
set upper limits on the contributions due to compound nuclear
effects and multi-step reactions at $<$1\%.
Our final
adopted ANC is
$C_{d_{5/2}}^2 = 1.08 \pm 0.10$ fm$^{-1}$ for the ground state.  The
uncertainty includes $\pm$4.8\% from the absolute normalization and
angle accuracies, plus the statistics of the fits, and $\pm$7.6\%
associated with the choice of optical model parameters and single
particle orbital, as well as ambiguities in the reaction mechanism. 
Our final adopted ANC is $C_{s_{1/2}}^2 = 6490 \pm
680$ fm$^{-1}$ for the first excited state.  The corresponding
contributions
to its uncertainty are $\pm$5.4\% and $\pm$9.0\%.

\section{S-Factors for $^{16}O(\lowercase{p},\gamma){}^{17}F$}

The relation of the ANC's to the direct capture rate at low
energies is
straightforward \cite{xu94}.  The cross section for the direct
capture
reaction $A + p \to B + \gamma$ can be written as 
\begin{equation}
\sigma=\lambda {\mid <I^{B}_{Ap}({\rm {\bf r}})\mid   
{\hat O}({\rm {\bf r}}) \mid \psi^{(+)}_{i}
({\rm {\bf r}})>
\mid}^2 ,
\label{capture}
\end{equation}
where $\lambda$ contains kinematic factors,
$I^{B}_{Ap}$ is the overlap
function for $B \to A + p$, ${\hat O}$ is the
electromagnetic transition operator, and
$\psi^{(+)}_{i}$ is the incident scattering wave.  If the
dominant contribution to
the matrix element comes from outside the nuclear radius, the
overlap function may be replaced by
\begin{equation}
I^{B}_{Ap}(r) \approx
C \frac{W_{-\eta,l+1/2}(2 \kappa r)}{r} ,
\label{approxeq}
\end{equation}
where $C$ defines the amplitude of the
tail of the radial
overlap function $I^{B}_{Ap}$, $W$ is the
Whittaker function, $\eta$ is the Coulomb
parameter for the bound state $B=A + p$, and
$\kappa$ is the bound state wave number.  For
${}^{16}O(p,\gamma){}^{17}F$, the necessary
$C$'s are just the ANC's determined from the $^{16}O(^3He,d)^{17}F$
transfer reaction studies in Sect.~II.  Thus, the direct capture
cross sections are directly proportional
to the squares of these ANC's.  In fact, the
${}^{16}O(p,\gamma){}^{17}F$
reaction populating the very weakly bound
$^{17}F$ first excited state provides
an extreme test of the connection
between the ANC measured in a transfer
reaction and the S-factor measured in direct capture.
The approximation of Eq.~(\ref{approxeq})
is excellent at large radii, but the
proximity of the node in the $2s_{1/2}$ wave
function makes it rather poor near the nuclear
surface.   In contrast, Eq.~(\ref{approxeq})
provides a good description of the
$^{17}F$ ground state
$1d_{5/2}$ wave function even in the vicinity
of the nuclear surface.

\begin{figure}
\begin{center}
  \mbox{\epsfysize=7.5cm \epsfbox{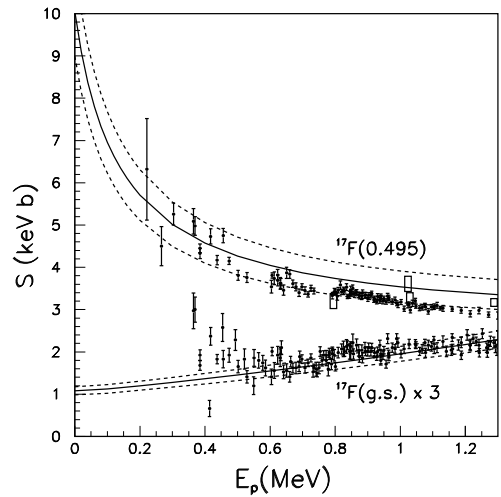}}
\end{center}
\caption{A comparison of the experimental
S-factors to those determined from
the ANC's found in
${}^{16}O({}^{3}He,d) {}^{17}F$.  The
solid data points are from
[9], and the open boxes are from [10].  The
solid lines indicate
our calculated S-factors, and the dashed lines
indicate the $\pm 1 \sigma$ error bands.  Note
that the experimental ground state S-factor
may be contaminated by background at energies
below 500 keV [25].}
\label{fig2}
\end{figure}

Following the prescription outlined above, the 
S-factors for
${}^{16}O(p,\gamma){}^{17}F$ were calculated with no free
parameters.
The results are shown in Fig.~\ref{fig2}.
Both $E1$ and $E2$ contributions have
been included in the calculations, but the $E1$
components dominate.
The capture of protons by $^{16}O$ at low
energies occurs at very large
distances $r$ due to the extremely small proton
separation
energy of $^{17}F$ \cite{capt1}.  Thus, we find
that
the calculated capture cross sections are sensitive
neither to the behavior of the overlap functions at
small $r$, nor to the nuclear
interaction between $^{16}O$ and $p$ in the initial
state \cite{capt2}.  We find that S(0) = 0.40 $\pm$ 0.04 keV$\cdot$b
for populating the $^{17}F$ ground state and S(0) = 9.8 $\pm$ 1.0
keV$\cdot$b for populating the first excited state.  The uncertainties
in these calculated zero-energy S-factors come almost entirely from
those in the corresponding ANC's determined above.  There is no
uncertainty associated with ambiguities in an extrapolation from
higher incident energies to zero energy, and there is very little
theoretical uncertainty, since the capture reaction is almost purely
peripheral at very low incident energies.
In the
astrophysical
domain, the energy dependence of the capture cross
sections is determined entirely by the
initial Coulomb scattering wave functions and the
kinematic factors,
while their magnitudes are fixed by the ANC's.
The theoretical uncertainty in the S-factors is
less than 2\% at an energy of 1 MeV.  This was estimated by
repeating the calculation while completely neglecting the nuclear
interaction in the initial state.  Hence, the
uncertainty in S at small
energies is due just to the uncertainties in the
ANC's measured above.
However, as the energy increases above 1 MeV, the calculated
S-factors become more sensitive to
the behavior of the overlap functions at smaller
$r$ and to the details of the nuclear interaction in the initial
state.  In that
case, the simple direct radiative
capture model used here breaks down, and a
microscopic approach including antisymmetrization is needed. 
This effect has been studied for
$^7Be(p,\gamma){}^8B$ in \cite{Csot97}.

Two previous measurements of
${}^{16}O(p,\gamma){}^{17}F$ have determined
the capture cross sections to the ground and first
excited states separately
\cite{capt1,capt2}.  The experimental results
for the S-factors populating the $^{17}F$
ground and excited states are also shown in
Fig.~\ref{fig2}.  It is clear from
Fig.~\ref{fig2} that the agreement between the
experimental results
and the predictions based on our measured
ANC's is indeed very good for proton energies below 1
MeV.  At these energies, the $^{16}O(p,\gamma){}^{17}F$ S-factors
derived
from the analysis of our
$^{16}O(^3He,d){}^{17}F$ measurements agree
with the corresponding direct experimental results
to better than 9\%.

Our calculated S-factors for $^{16}O(p,\gamma){}^{17}F$ in
Fig.~\ref{fig2} are very similar to the S-factors calculated for the
same reaction in \cite{capt1}.  The energy dependences are virtually
identical.  For both states, we calculate the S-factor to be slightly
larger than those in \cite{capt1}, which provides us with a somewhat
better representation of the ground state S-factor and a slightly
poorer representation for the first excited state.  It is important
to recognize that the procedures used in the two calculations are
very different, even though their final results are quite similar.
In \cite{capt1}, the ${}^{17}F$ ground and first excited states
were assumed to be good single particle states outside a closed
$^{16}O$
core.  Electron scattering data were used to specify the
density distribution of $^{16}O$, which provided the input for a
folding model calculation of
the low energy $p - {}^{16}O$ potential with DDM3Y.  The central
and spin-orbit terms in the potential were renormalized separately,
for both
even and odd partial waves, by fitting the ${}^{17}F$ bound state
energies and comparing to detailed data on low energy
$p + {}^{16}O$ elastic scattering.  Finally, the direct
capture rates were calculated with no additional free parameters. 
This level of
detail was necessary to reproduce the $^{16}O(p,\gamma){}^{17}F$
S-factors at proton
energies higher than we consider here.

The ANC technique is quite different, much simpler, and based on
much less experimental input.  Our measured $^{16}O(^3He,d){}^{17}F$
angular distributions determined the ANC's for $^{17}F \to {}^{16}O +
p$ experimentally.  These specify the amplitudes of the tails of the
$^{17}F \to {}^{16}O + p$ overlap functions.  We then normalized
single particle orbitals to the measured ANC's, and used them to
calculate the corresponding direct radiative capture S-factors.  So
long as one restricts oneself to the low energies typically of
greatest importance to nuclear astrophysics, the only input required
by this technique is the experimentally measured value of the ANC. 
In practice, the close agreement between our calculated S-factors and
those in \cite{capt1} indicate that the body of experimental data
used to specify the $p + {}^{16}O$ potential in \cite{capt1}
ultimately was sufficient to determine the $^{17}F$ ANC's indirectly.
However, the ANC approach may also be used to determine S-factors for
direct radiative capture from peripheral transfer reaction data in
cases, such as radioactive targets,
for which much less experimental data are available than for
$^{16}O$.

\section{Conclusion}

In conclusion, the
$^{16}O(p,\gamma){}^{17}F$ S-factors derived
from the analysis of our
$^{16}O(^3He,d){}^{17}F$ measurements agree
with the corresponding direct experimental results
to better than 9\%.  This demonstrates the
practicality of determining accurate S-factors
for very low energy direct capture reactions from
measurements of the corresponding
asymptotic normalization coefficients in
peripheral proton transfer reactions.   This
technique can be extended to other systems,
including
measurements with radioactive beams.  The
production of ${}^{8}B$ in the sun
via the  ${}^{7}Be (p,\gamma) {}^{8}B$
reaction is an ideal example.  While this
reaction is relatively unimportant in the
production of energy, it
provides the only source of high energy
neutrinos.  Hence its rate is
crucial to interpreting measurements from
solar neutrino detectors \cite{rmp98}.  At
stellar energies this reaction is completely
dominated by direct capture
which occurs at large radii.  Indeed, even
before this demonstration of the accuracy
of this indirect technique, there has been an
attempt \cite{Liu96} to
determine the $^7Be(p,\gamma){}^8B$ S-factor
from a measurement of the
$^8B \to {}^7Be + p$ ANC in the reaction
$^2H(^7Be,{}^8B)n$.  But interpretation
of that result suffered from significant
uncertainties in the choice of optical potentials
\cite{Gagl98}, at least in part due
to the very low energies involved.
The $^8B \to {}^7Be + p$ ANC can also be
measured in $({}^{7}Be,{}^{8}B)$ transfer
reactions at higher energies
with heavier targets, where the uncertainties
due to the choice of optical
potentials are much reduced.  We will report cross sections for
this reaction using ${}^{10}B$ and ${}^{14}N$
targets in future publications.

\acknowledgments

This work was supported in part by the U.S.
Department of Energy under
Grant number
DE-FG05-93ER40773 and by the Robert A.
Welch Foundation.

\end{document}